\documentclass[runningheads]{llncs}
\usepackage{graphicx}
\usepackage{booktabs}
\usepackage{multirow}
\usepackage{float}
\usepackage[colorlinks=true,hyperfootnotes=false]{hyperref}
\usepackage{graphicx}

\usepackage{amsthm}
\usepackage{amsmath}
\usepackage{amssymb}
\usepackage{mathrsfs}
\usepackage{algorithm}
\usepackage{algorithmic}

\usepackage{color}
\usepackage{xcolor}

\newcommand{\TheName}{\texttt{LightM-UNet}}
\begin{document}

\title{\TheName{}: Mamba Assists in Lightweight UNet for Medical Image Segmentation}

% \author{Anonymous}
% \institute{Anonymous Organization \\
% \email{**@***.**}}

\author{
Weibin Liao$^{1,3}$\and
Yinghao Zhu$^{2,4}$\and
Xinyuan Wang$^{4}$\and
Chengwei Pan$^{4}$\and
Yasha Wang$^{1,2~\dagger}$\and
Liantao Ma$^{1,2~\dagger}$
}
\institute{
$^1$Key Laboratory of High Confidence Software Technologies, Ministry of Education, Beijing, China\\
$^2$National Engineering Research Center for Software Engineering, Peking University, Beijing, China\\
$^3$School of Computer Science, Peking University, Beijing, China\\
$^4$Institute of Artificial Intelligence, Beihang University, Beijing, China
}

\authorrunning{W. Liao et al.}
\titlerunning{LightM-UNet}

\maketitle              % typeset the header of the contribution

\begin{abstract}
UNet and its variants have been widely used in medical image segmentation. However, these models, especially those based on Transformer architectures, pose challenges due to their large number of parameters and computational loads, making them unsuitable for mobile health applications.
Recently, State Space Models (SSMs), exemplified by Mamba, have emerged as competitive alternatives to CNN and Transformer architectures. 
Building upon this, we employ Mamba as a lightweight substitute for CNN and Transformer within UNet, aiming at tackling challenges stemming from computational resource limitations in real medical settings.
To this end, we introduce the Lightweight Mamba UNet (\TheName{}) that integrates Mamba and UNet in a lightweight framework. Specifically, \TheName{} leverages the \textit{Residual Vision Mamba Layer} in a pure Mamba fashion to extract deep semantic features and model long-range spatial dependencies, with linear computational complexity.
Extensive experiments conducted on two real-world 2D/3D datasets demonstrate that \TheName{} surpasses existing state-of-the-art literature. Notably, when compared to the renowned nnU-Net, \TheName{} achieves superior segmentation performance while drastically reducing parameter and computation costs by 116x and 21x, respectively. 
This highlights the potential of Mamba in facilitating model lightweighting.
Our code implementation is publicly available at \url{https://github.com/MrBlankness/LightM-UNet}

\end{abstract}

\vspace{-6mm}
\section{Introduction}

\begin{figure}[!t]
\centering
\includegraphics[width=\textwidth]{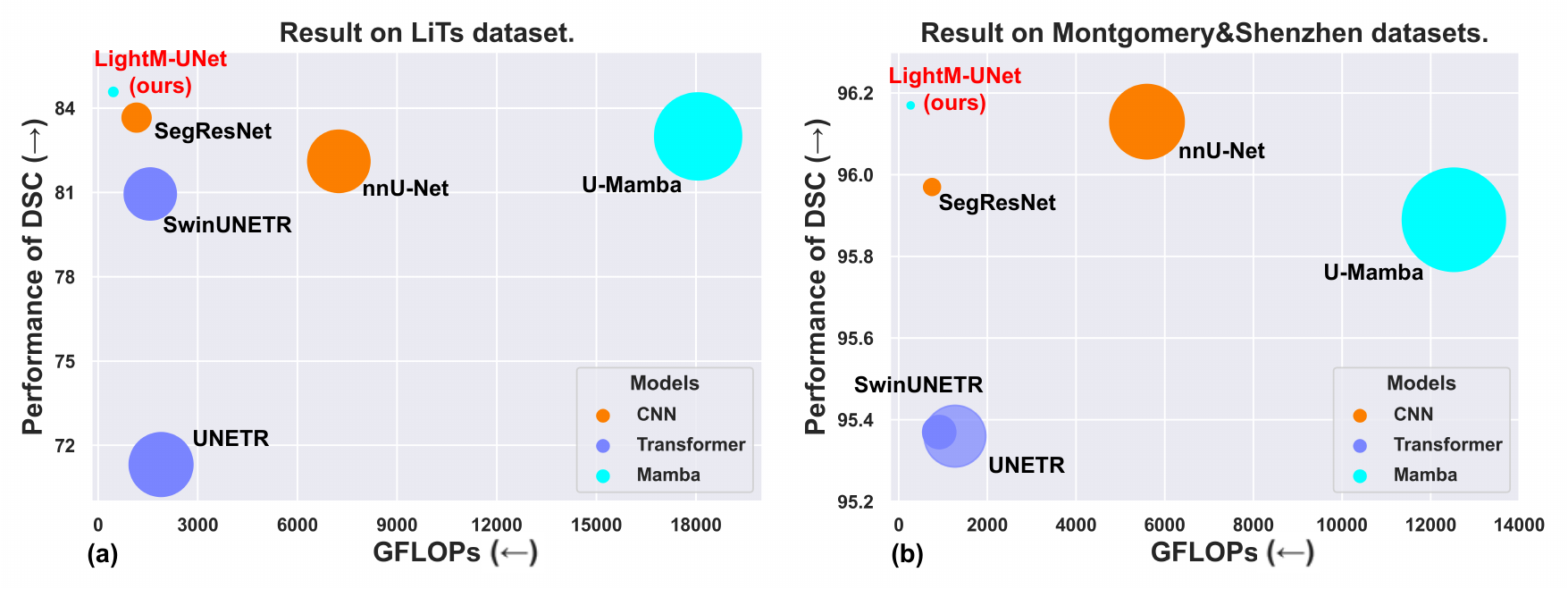}
\vspace{-3mm}
\caption{(a) and (b) respectively show the visualization of comparative experimental results on LiTs~\cite{bilic2023liver} and Montgomery\&Shenzhen~\cite{jaeger2014two} datasets. 
The central position of the marker indicates the performance of the model, while the size of the marker indicates the number of parameters of the model (larger size indicates a greater number of parameters). Colors in the legend represent the basic architecture these models applied.} \label{fig:main_result}
\vspace{-3mm}
\end{figure}

UNet~\cite{ronneberger2015u}, as a well-established algorithm for medical image segmentation, finds extensive application across a spectrum of segmentation tasks pertaining to medical organs and lesions, spanning various modalities of medical images. Its symmetrical U-shaped encoder-decoder architecture, coupled with integral skip connections, has laid the groundwork for segmentation models, spawning a plethora of works~\cite{isensee2021nnu,myronenko20193d,ruan2023ege} predicated on the U-shaped structure.
However, being a Convolutional Neural Network-based (CNN-based) model, UNet grapples with the inherent locality of convolution operations, which poses limitations on its capacity to apprehend explicit global and long-range semantic information interactions~\cite{chen2021transunet}. Several studies have attempted to mitigate this issue by employing atrous convolutional layers~\cite{gu2019net}, self-attention mechanisms~\cite{schlemper2019attention}, and image pyramids~\cite{zhao2017pyramid}. Nonetheless, these methods still exhibit constraints in modeling long-range dependencies.

In efforts to endow UNet with the capacity to apprehend global information, recent studies~\cite{chen2021transunet,hatamizadeh2022unetr,hatamizadeh2021swin} have delved into integrating Transformer architectures~\cite{vaswani2017attention}, leveraging self-attention mechanisms to capture global information by treating the image as a sequence of contiguous patches. Although effective, Transformer-based solutions introduce quadratic complexity concerning image sizes owing to the self-attention mechanism, leading to a substantial computational overhead, particularly for tasks requiring dense predictions such as medical image segmentation.
This overlooks the imperative of computational constraints in real-world medical settings, falling short of fulfilling the necessity for models characterized by low parameters and minimal computational load in mobile healthcare segmentation tasks~\cite{ruan2023ege}.
In summary, the unresolved inquiry persists: \textit{``How can UNet be endowed with the capability to accommodate long-range dependencies without incurring additional parameters and computational burden?''}

Recently, State Space Models (SSMs) have garnered considerable attention among researchers. Expanding upon the groundwork laid by classical SSM research~\cite{kalman1960new}, modern SSMs (e.g., Mamba~\cite{gu2023mamba}) not only establish long-range dependencies but also demonstrate linear complexity concerning input size, making Mamba a strong competitor to CNN and Transformer on the lightweight road of UNet.
Some contemporary endeavors, such as U-Mamba~\cite{ma2024u}, have proposed a hybrid CNN-SSM block, amalgamating the local feature extraction capability of convolutional layers with SSM's proficiency in capturing longitudinal dependency relationships. However, U-Mamba~\cite{ma2024u} introduces a substantial number of parameters and computational load (173.53M parameters and 18,057.20 GFLOPs), rendering it challenging to deploy in mobile healthcare settings for medical segmentation tasks.
Therefore, in this study, we introduce \TheName{}, a lightweight U-shaped segmentation model based on Mamba, which achieves state-of-the-art performance while significantly reducing parameter and computation costs (as depicted in Fig.~\ref{fig:main_result}). The contributions of this work are threefold.

\begin{enumerate}
    \item We introduce \TheName{}, a lightweight fusion of UNet and Mamba, boasting a mere parameter count of \textbf{1M}. Through validation on both 2D and 3D real-world datasets, \TheName{} surpasses existing state-of-the-art models. In comparison to the renowned nnU-Net~\cite{isensee2021nnu} and contemporaneous U-Mamba~\cite{ma2024u}, \TheName{} reduces the parameter count by 116$\times$ and 224$\times$, respectively.
    \item Technically, we propose the \textit{Residual Vision Mamba Layer (RVM Layer)} to extract deep features from images in a pure Mamba manner. With minimal introduction of new parameters and computational overhead, we further enhance the capability of SSM to model long-range spatial dependencies in visual images by utilizing \textit{residual connections} and \textit{adjustment factors}.
    \item Insightly, in contrast to contemporaneous endeavors~\cite{ma2024u,xing2024segmamba,ruan2024vm} that integrate UNet with Mamba, we advocate for employing Mamba as a lightweight substitute for CNN and Transformer within UNet, aiming at tackling challenges stemming from computational resource limitations in real medical settings. To our knowledge, this represents the pioneering effort introducing Mamba into UNet as a lightweight optimization strategy.
\end{enumerate}

\section{Methodologies}

\hspace{0.5cm} \textit{While \TheName{} supports both 2D and 3D versions of medical image segmentation, for convenience, this manuscript describes the methodology using the 3D version of \TheName{}.} \hfill  \textit{\textbf{ --- A reading-friendly reminder.}}

\subsection{Architecture Overview}

\begin{figure}[!t]
\centering
\includegraphics[width=0.8\textwidth]{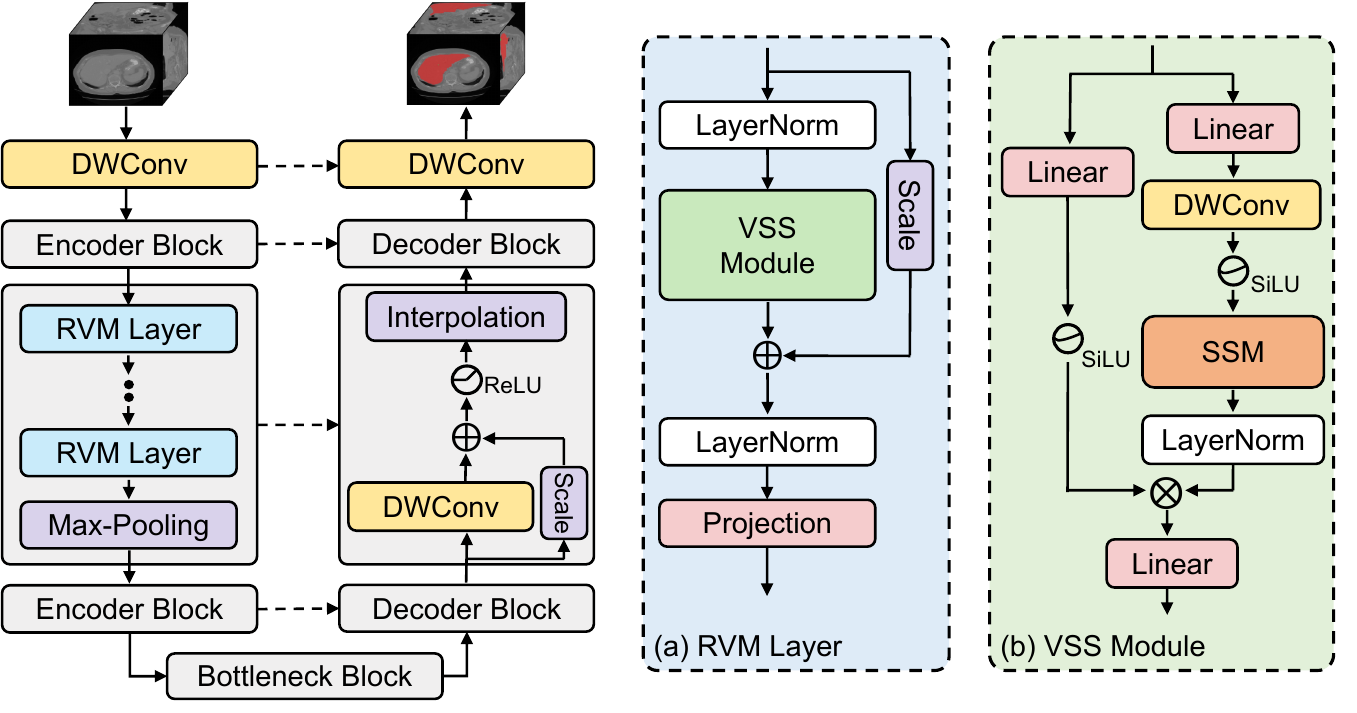}
\vspace{-1mm}
\caption{The overall network architecture of \TheName{} as well as the (a) \textit{Residual Vision Mamba Layer (RVM Layer)}, the (b) \textit{Vision State-Space Module (VSS Module)}.} \label{fig:framework}
\vspace{-3mm}
\end{figure}

The overall architecture of the proposed \TheName{} is illustrated in Fig.~\ref{fig:framework}. 
Given an input image $I \in \mathbb{R}^{C \times H \times W \times D}$, where $C$, $H$, $W$, and $D$ denote the number of channels, height, width, and number of slices of the 3D medical image, respectively.
\TheName{} commences by employing a depthwise convolution (DWConv) layer for shallow feature extraction, generating the shallow feature map $F_S \in \mathbb{R}^{32 \times H \times W \times D}$, where 32 denotes a fixed number of filters. 
Subsequently, \TheName{} incorporates three consecutive Encoder Blocks to extract deep features from the images. 
Post each Encoder Block, the number of channels in the feature maps doubles, while the resolution halves. 
Consequently, \TheName{} extracts deep features $F_{D}^{l} \in \mathbb{R}^{(32 \times 2^l) \times (H / 2^l) \times (W / 2^l) \times (D / 2^l)}$ at the $l$-th Encoder Block, where $l \in \{1, 2, 3\}$.
Subsequent to this, \TheName{} employs a Bottleneck Block to model long-range spatial dependencies while retaining the size of the feature maps unchanged. 
Following that, \TheName{} integrates three consecutive Decoder Blocks for feature decoding and image resolution restoration. 
Following each Decoder Block, the number of channels in the feature maps is halved, and the resolution is doubled.
Finally, the output of the last Decoder Block attains the same resolution as the original image, comprising 32 feature channels. 
\TheName{} utilizes a DWConv layer to map the number of channels to the number of segmentation targets and applies a SoftMax activation function to generate the image mask. 
In alignment with the design of UNet, \TheName{} also employs skip connections to furnish multi-level feature maps for the decoder.

\subsection{Encoder Block}

To minimize the number of parameters and computational cost, \TheName{} employs Encoder Blocks comprising solely Mamba structures to extract deep features from the image. Specifically, given a feature map $F^{l} \in \mathbb{R}^{\check{C} \times \check{H} \times \check{W} \times \check{D}}$, where $\check{C} = 32 \times 2^l$, $\check{H} = H / 2^l$, $\check{W} = W / 2^l$, $\check{D} = D / 2^l$, and $l \in \{1, 2, 3\}$, the Encoder Block initially flattens and transposes the feature map into a shape of $(\check{L}, \check{C})$, where $\check{L} = \check{H} \times \check{W} \times \check{D}$.
Subsequently, the Encoder Block utilizes $N_l$ consecutive RVM Layers to capture global information, with the number of channels increased in the last RVM Layer. Following this, the Encoder Block reshapes and transposes the feature map into a shape of $(\check{C} \times 2, \check{H}, \check{W}, \check{D})$, succeeded by a Max-Pooling operation to reduce the resolution of the feature map. Ultimately, the $l$-th Encoder Block outputs the new feature map $F^{l+1}$ with a shape of $(\check{C} \times 2, \check{H} / 2, \check{W} / 2, \check{D} / 2)$.

\paragraph{\textbf{Residual Vision Mamba Layer (RVM Layer)}}

\TheName{} proposes the RVM Layer to enhance the original SSM block for image deep semantic feature extraction. 
Specifically, \TheName{} utilizes advanced \textit{residual connections} and \textit{adjustment factors} to further enhance the long-range spatial modeling capability of SSM, with almost no introduction of new parameters and computational complexity.
As depicted in Fig.~\ref{fig:framework}~(a), given the input deep feature $M_{in}^{l} \in \mathbb{R}^{\check{L} \times \check{C}}$, the RVM Layer initially employs LayerNorm followed by the VSSM to capture spatial long-range dependencies. Subsequently, it utilizes an adjustment factor $s \in \mathbb{R}^{\check{C}}$ in the residual connection\cite{chen2024recursive} for better performance.
This process can be represented mathematically as follows:

\begin{equation}
    \widetilde{M^l} = VSSM(LayerNorm(M_{in}^{l})) + s \cdot M_{in}^{l}
\end{equation}

Following this, the RVM Layer employs another LayerNorm to normalize $\widetilde{M^l}$, and subsequently utilizes a projection layer to convert $\widetilde{M^l}$ into a deeper feature.
The above process can be formulated as:
\begin{equation}
    M_{out}^l = Projection(LayerNorm(\widetilde{M^l}))
\end{equation}

\paragraph{\textbf{Vision State-Space Module (VSS Module)}}

Following the approach outlined in \cite{liu2024vmamba}, \TheName{} introduces the VSS Module (depicted in Fig.~\ref{fig:framework}~(b)) for long-range spatial modeling. The VSS Module takes the feature $W_{in}^{l} \in \mathbb{R}^{\check{L} \times \check{C}}$ as input and channels it into two parallel branches.
In the first branch, the VSS Module expands the feature channels to $\lambda \times \check{C}$ using a linear layer, where $\lambda$ represents a pre-defined channel expansion factor. Subsequently, it applies a DWConv, SiLU activation function~\cite{shazeer2020glu}, followed by the SSM and LayerNorm.
In the second branch, the VSS Module also expands the feature channels to $\lambda \times \check{C}$ using a linear layer, followed by the SiLU activation function.
Subsequently, the VSS Module aggregates features from the two branches using the Hadamard product and projects the channel number back to $\check{C}$ to generate the output $W_{out}$ with the same shape as the input $W_{in}$.
The above process can be formulated as:
\begin{equation}
\begin{gathered}
    W_1 = LayerNorm(SSM(SiLU(DWConv(Linear(W_{in}))))) \\
    W_2 = SiLU(Linear(W_{in})) \\
    W_{out} = Linear(W_1 \odot W_2)
\end{gathered}
\end{equation}
where $\odot$ denotes the Hadamard product.

\vspace{-3mm}
\subsection{Bottleneck Block}

Similar to Transformer, Mamba encounters convergence challenges when the network depth becomes excessive~\cite{touvron2021going}. Consequently, \TheName{} addresses this issue by incorporating four successive RVM Layers to construct bottlenecks for further modeling spatial long-term dependency. Within these bottleneck regions, the number of feature channels and the resolution remain unchanged.

\vspace{-3mm}
\subsection{Decoder Block}

\TheName{} employs Decoder Blocks to decode feature maps and restore image resolution. Specifically, given $F_D^l \in \mathbb{R}^{\check{C} \times \check{H} \times \check{W} \times \check{D}}$ from the skip connection and $P_{in} \in \mathbb{R}^{\check{C} \times \check{H} \times \check{W} \times \check{D}}$ from the output of the previous block, the Decoder Block first performs feature fusion using an addition operation. Subsequently, it utilizes a DWConv, a residual connection, and a ReLU activation function to decode the feature map. Additionally, an \textit{adjustment factors} $s'$ is added to the residual connection to enhance the decoding capability.
This process can be expressed mathematically as:
\begin{equation}
    P_{out} = ReLU(DWConv(P_{in} + F_D^l) + s' \cdot (P_{in} + F_D^l))
\end{equation}
The Decoder Block ultimately employs bilinear interpolation to restore predictions to the original resolution.

\section{Experiments}
\paragraph{\textbf{Datasets and Experimental Setups.}}
To assess the performance of our model, we select two publicly available medical image datasets: the LiTs dataset~\cite{bilic2023liver}, comprising 3D CT images, and the Montgomery\&Shenzhen dataset~\cite{jaeger2014two}, comprising 2D X-ray images. These datasets are extensively utilized in existing segmentation research~\cite{liao2022muscle,zhang2021multi} and are employed here to validate the performance of the 2D and 3D versions of \TheName{}, respectively. The data were randomly partitioned into training, validation, and testing sets in a ratio of 7:1:2.

\TheName{} was implemented using the PyTorch framework and the number of RVM Layers in the three Encoder Blocks is set as 1, 2, and 2, respectively. All experiments were conducted on a single Quadro RTX 8000 GPU. SGD was employed as the optimizer, initialized with a learning rate of 1e-4. The PolyLRScheduler was used as the scheduler, and a total of 100 epochs were trained. In addition, the loss function was designed as a simple combination of Cross Entropy loss and Dice loss.
For the LiTs dataset, the images were normalized and resized to 128 $\times$ 128 $\times$ 128, with a batch size of 2. For the Montgomery\&Shenzhen dataset~\cite{jaeger2014two}, the images were normalized and resized to 512 $\times$ 512, with a batch size of 12.

To evaluate \TheName{}, we compared it with two CNN-based segmentation networks (nnU-Net~\cite{isensee2021nnu} and SegResNet~\cite{myronenko20193d}), two Transformer-based networks (UNETR~\cite{hatamizadeh2022unetr} and SwinUNETR~\cite{hatamizadeh2021swin}), and a Mamba-based network (U-Mamba \cite{ma2024u}), which are commonly used in medical image segmentation competitions. Additionally, we employed Mean Intersection over Union (mIoU) and Dice similarity score (DSC) as evaluation metrics.

\begin{table}[!t]
\centering
\caption{Comparative experimental results on the LiTS~\cite{bilic2023liver} dataset using various 3D segmentation models.}
\label{tab:LiTs_main_result}
\begin{tabular}{lrrrrrrrr}
\toprule
\multicolumn{1}{c}{\multirow{2}{*}{\textbf{Models}}} & \multicolumn{1}{c}{\multirow{2}{*}{\textbf{Params(M)}}} & \multicolumn{1}{c}{\multirow{2}{*}{\textbf{GFLOPs}}} & \multicolumn{2}{c}{\textbf{Liver}} & \multicolumn{2}{c}{\textbf{Tumor}} & \multicolumn{2}{c}{\textbf{Average}} \\
\cmidrule(r){4-5} \cmidrule(r){6-7} \cmidrule(r){8-9}
\multicolumn{1}{c}{} & \multicolumn{1}{c}{} & \multicolumn{1}{c}{} & \multicolumn{1}{c}{\textbf{DSC}} & \multicolumn{1}{c}{\textbf{mIoU}} & \multicolumn{1}{c}{\textbf{DSC}} & \multicolumn{1}{c}{\textbf{mIoU}} & \multicolumn{1}{c}{\textbf{DSC}} & \multicolumn{1}{c}{\textbf{mIoU}} \\
\midrule
nnU-Net~\cite{isensee2021nnu} & 88.62 & 7,240.26 & 95.77 & 91.94 & 68.45 & 56.34 & 82.11 & 74.13 \\
SegResNet~\cite{myronenko20193d} & 18.79 & 1,158.30 & 96.11 & 92.56 & 71.22 & 59.76 & 83.67 & 76.16 \\
UNETR~\cite{hatamizadeh2022unetr} & 92.62 & 1,891.35 & 94.06 & 88.95 & 48.58 & 37.01 & 71.32 & 62.98 \\
SwinUNETR~\cite{hatamizadeh2021swin} & 61.99 & 1,570.32 & 95.24 & 91.07 & 66.67 & 55.09 & 80.95 & 73.08 \\
U-Mamba~\cite{ma2024u} & 173.53 & 18,057.20 & 95.94 & 92.33 & 70.05 & 58.42 & 83.00 & 75.37 \\
\TheName{} & \textbf{1.87} & \textbf{457.62} & \textbf{96.31} & \textbf{92.92} & \textbf{72.86} & \textbf{62.05} & \textbf{84.58} & \textbf{77.48} \\
\bottomrule
\end{tabular}
\vspace{-2mm}
\end{table}

\begin{table}[!t]
\centering
\caption{Comparative experimental results on the Montgomery\&Shenzhen~\cite{jaeger2014two} dataset using various 2D segmentation models. }
\label{tab:MS_main_result}
\begin{tabular}{lrrrr}
\toprule
\multicolumn{1}{c}{\textbf{Models}} & \multicolumn{1}{c}{\textbf{Params(M)}} & \multicolumn{1}{c}{\textbf{GFLOPs}} & \multicolumn{1}{c}{\textbf{DSC}} & \multicolumn{1}{c}{\textbf{mIoU}} \\
\midrule
nnU-Net~\cite{isensee2021nnu} & 126.56 & 5,594.98 & 96.13 & 92.66 \\
SegResNet~\cite{myronenko20193d} & 6.29 & 748.96 & 95.97 & 92.36 \\
UNETR~\cite{hatamizadeh2022unetr} & 87.51 & 1,267.53 & 95.36 & 91.26 \\
SwinUNETR~\cite{hatamizadeh2021swin} & 25.12 & 909.26 & 95.37 & 91.31 \\
U-Mamba~\cite{ma2024u} & 244.10 & 12,521.27 & 95.89 & 92.23 \\
\TheName{} & \textbf{1.09} & \textbf{267.19} & \textbf{96.17} & \textbf{92.74} \\
\bottomrule
\end{tabular}
\vspace{-3mm}
\end{table}

\paragraph{\textbf{Comparative Results.}}

The comparative experimental results presented in Table.~\ref{tab:LiTs_main_result} demonstrate that our \TheName{} achieves comprehensive state-of-the-art performance on the LiTS dataset\cite{li2023efficient}. Notably, compared to larger models like nnU-Net, \TheName{} not only exhibits superior performance but also significantly reduces the number of parameters and computational costs by 47.39$\times$ and 15.82$\times$, respectively.
When compared to the contemporaneous U-Mamba~\cite{ma2024u}, \TheName{} shows a performance improvement of 2.11\% in terms of average mIoU. Particularly for tumors, which are often too small to be easily detected, \TheName{} achieves a mIoU improvement of 3.63\%. Importantly, as a method incorporating Mamba into the UNet architecture, \TheName{} utilizes only 1.07\% fewer parameters and 2.53\% less computational resources compared to U-Mamba~\cite{ma2024u}.

\begin{figure}[!t]
\centering
\includegraphics[width=\textwidth]{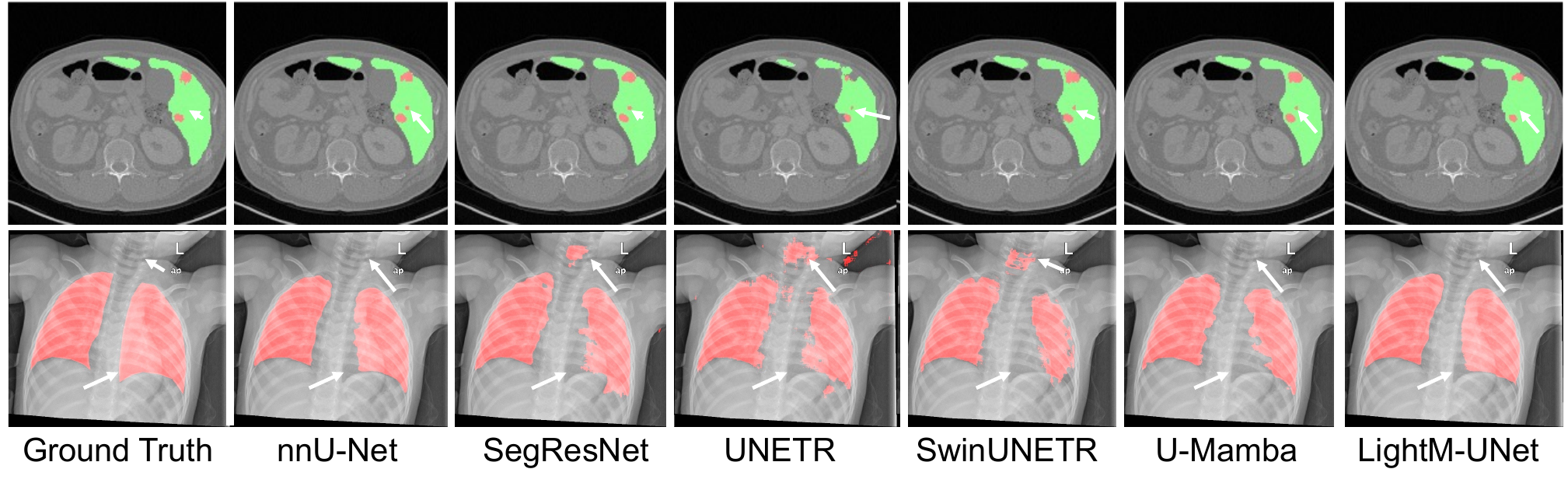}
\vspace{-7mm}
\caption{Visualized segmentation examples of LiTS \cite{bilic2023liver} (1st row, \textbf{\textcolor{red}{red}} parts indicate tumor and \textbf{\textcolor{green}{green}} parts indicate liver) and Montgomery\&Shenzhen \cite{jaeger2014two} (2nd row, \textbf{\textcolor{red}{red}} parts indicate lung) datasets. The white arrows point to the parts where significant differences exist in various segmentation results.} \label{fig:segmentation_visual}
\vspace{-1mm}
\end{figure}

The experimental results for the Montgomery\&Shenzhen datasets~\cite{jaeger2014two} are summarized in Table.~\ref{tab:MS_main_result}. 
\TheName{} once again achieves optimal performance and significantly surpassed other Transformer-based and Mamba-based literature.
Besides, \TheName{} stands out for its remarkably low parameter count, utilizing only 1.09M parameters. This represents a reduction in parameters by 99.14\% and 99.55\% compared to nnU-Net~\cite{isensee2021nnu} and U-Mamba~\cite{ma2024u}, respectively.
For a more clear visualization of the experimental findings, please refer to Fig.~\ref{fig:main_result}. 
Fig.~\ref{fig:segmentation_visual} illustrates segmentation result examples demonstrating that compared to other models, \TheName{} exhibits smoother segmentation edges and does not produce erroneous identification for small objects (such as tumors).

\vspace{-3mm}
\paragraph{\textbf{Ablation Results.}} 

We conduct extensive ablation experiments to demonstrate the effectiveness of our proposed modules. 
We first analyze the performance of CNN, Transformer, and Mamba within the UNet framework. Specifically, we replace the VSS Module in \TheName{} with a convolution operation with a $3 \times 3$ kernel for CNN and with the self-attention mechanism for the Transformer. Considering the memory constraints, for CNN, we replace all VSS Modules in \TheName{}, while for the Transformer, we follow the design of TransUNet~\cite{chen2021transunet} and only replace the VSS Modules in the Bottleneck Block.
The experimental results on LiTS dataset~\cite{bilic2023liver} are shown in Table.~\ref{tab:ablation_study}, indicating that replacing VSSM with either Convolution or Self-Attention leads to performance sacrifices. Additionally, Convolution and Self-Attention introduces a significant number of parameters and computational overhead. Furthermore, we observe that both Transformer-based and VSSM-based results outperform Convolution-based results, demonstrating the benefits of modeling long-range dependencies.

We further remove the \textit{Adjustment factors} and \textit{Residual connections} in the RVM Layer. Experimental results show that after removing these two components, the model's parameter count and computational overhead hardly decrease, but the model's performance significantly declines (0.44\%$\downarrow$ and 0.69\%$\downarrow$ mIoU). This validates our basic principle of enhancing model performance without introducing additional parameters and computational overhead.
The additional ablation analysis on the Montgomery\&Shenzhen datasets~\cite{jaeger2014two} can be found in the supplementary material.

\begin{table}[!t]
\centering
\caption{Ablation studies in Residual Vision Mamba Layer on the LiTS~\cite{bilic2023liver} dataset.}
\label{tab:ablation_study}
\resizebox{\columnwidth}{!}{
\begin{tabular}{lrrrrrrrr}
\toprule
\multicolumn{1}{c}{\multirow{2}{*}{\textbf{Models}}} & \multicolumn{1}{c}{\multirow{2}{*}{\textbf{Params(M)}}} & \multicolumn{1}{c}{\multirow{2}{*}{\textbf{GFLOPs}}} & \multicolumn{2}{c}{\textbf{Liver}} & \multicolumn{2}{c}{\textbf{Tumor}} & \multicolumn{2}{c}{\textbf{Average}} \\
\cmidrule(r){4-5} \cmidrule(r){6-7} \cmidrule(r){8-9}
\multicolumn{1}{c}{} & \multicolumn{1}{c}{} & \multicolumn{1}{c}{} & \multicolumn{1}{c}{\textbf{DSC}} & \multicolumn{1}{c}{\textbf{mIoU}} & \multicolumn{1}{c}{\textbf{DSC}} & \multicolumn{1}{c}{\textbf{mIoU}} & \multicolumn{1}{c}{\textbf{DSC}} & \multicolumn{1}{c}{\textbf{mIoU}} \\
\midrule
VSSM$\rightarrow$Conv3 & 18.79 & 1,513.44 & 96.11 & 92.56 & 71.22 & 59.76 & 83.67 & 76.16 \\
VSSM$\rightarrow$Self-Attention & 3.44 & 470.50 & 96.09 & 92.54 & 72.53 & 61.06 & 84.31 & 76.80 \\
\midrule
w/o Adjustment factors & 1.87 & 457.62 & 96.28 & 92.86 & 71.54 & 60.73 & 83.91 & 76.79 \\
w/o Residual connections & 1.87 & 457.62 & 96.22 & 92.76 & 72.53 & 61.32 & 84.38 & 77.04 \\
\midrule
\TheName{} & \textbf{1.87} & \textbf{457.62} & \textbf{96.31} & \textbf{92.92} & \textbf{72.86} & \textbf{62.05} & \textbf{84.58} & \textbf{77.48} \\
\bottomrule
\end{tabular}
}
\vspace{-5mm}
\end{table}

\vspace{-3mm}
\section{Conclusion}
\vspace{-2mm}

In this study, we introduce \TheName{}, a lightweight network based on Mamba, which achieves state-of-the-art performance in both 2D and 3D segmentation tasks while comprising only \textbf{1M} parameters, over 99\% lower parameters and significant lower GFLOPS against latest Transformer-based architectures. We validate our approach through rigorous ablation studies within a unified framework, marking the initial attempt to employ Mamba as a lightweight strategy for UNet.
Our future work includes designing a more lightweight network and validate on more datasets of multiple organs, fostering their applications in mobile health and beyond.

\bibliographystyle{splncs04}
\bibliography{main_arxiv}

\end{document}